\documentclass[aps,prd,showpacs,superscriptaddress,nofootinbib,notitlepage]{revtex4-1}
\usepackage{epsfig}
\newcommand{\ben}{\begin{eqnarray}}
\newcommand{\een}{\end{eqnarray}}
\newcommand{\nnu}{\nonumber\\}
\newcommand{\bef}{\begin{figure}[htb]\centering}
\newcommand{\eef}{\end{figure}}

\begin{document}
\title{QCD evolution of naive-time-reversal-odd parton distribution functions}

\author{Zhong-Bo Kang}
\email{zkang@bnl.gov}
\affiliation{Theoretical Division, 
                   Los Alamos National Laboratory, 
                   Los Alamos, NM 87545}
\affiliation{RIKEN BNL Research Center,
                   Brookhaven National Laboratory,
                   Upton, NY 11973}

\author{Jian-Wei Qiu}
\email{jqiu@bnl.gov}
\affiliation{Physics Department, 
                   Brookhaven National Laboratory, 
                   Upton, NY 11973}
\affiliation{C.N. Yang Institute for Theoretical Physics, 
                   Stony Brook University, 
                   Stony Brook, NY 11794}

\begin{abstract}
We reexamine the derivation of the leading order QCD evolution equations of twist-3 quark-gluon correlation functions, $T_{q,F}(x,x)$ and $T^{(\sigma)}_{q,F}(x,x)$, which are the first transverse-momentum-moment of the naive-time-reversal-odd parton distribution functions - the Sivers and Boer-Mulders function, respectively.  The evolution equations were derived by several groups with apparent differences.  We identify the sources that are responsible for the differences, and are able to reconcile the results from various groups. 
\end{abstract}

\date{\today}
\maketitle

Transverse spin physics has attracted tremendous attention from both experimental and theoretical communities in recent years \cite{D'Alesio:2007jt}. These effects typically manifest themselves in various azimuthal asymmetries. The well-known examples are the single transverse spin asymmetry in polarized proton-proton collisions \cite{SSA-rhic}, Sivers and Collins asymmetries in the semi-inclusive hadron production in deep inelastic scattering \cite{Qian:2011py}, the Boer-Mulders effect in Drell-Yan production \cite{Zhu:2006gx}, as well as the large $\cos(2\phi)$ anomalous azimuthal asymmetry in back-to-back dihadron production in $e^+e^-$ annihilation \cite{belle}. It was soon realized that these non-trivial azimuthal asymmetries in high energy collisions should be directly connected to the transverse motion of partons inside the parent hadron. Experimental measurements of the asymmetries and the investigation to understand the underlying dynamics have provided and will continue to provide us new opportunities to explore QCD and the hadron structure far beyond what we have been able to achieve.

Two complementary QCD-based approaches have been proposed to analyze the physics behind the measured asymmetries: the transverse momentum dependent (TMD) factorization approach \cite{TMD-fac,Brodsky,MulTanBoe,boermulders} and the collinear twist-three factorization approach \cite{Efremov,qiu,koike}. In the TMD factorization approach, the asymmetry was attributed to the spin and transverse momentum correlation between the identified hadron and the active parton, which are represented by the TMD parton distribution or fragmentation function.  On the other hand, in the collinear factorization approach, all active partons' transverse momenta are integrated into the collinear distributions, and the explicit correlation between the spin and the transverse momentum in the TMD approach is now included into the high twist collinear parton distributions or fragmentation functions.  The asymmetry in the collinear factorization approach is represented by twist-3 collinear parton distributions or fragmentation functions, which have no probability interpretation, and could be interpreted as the quantum interference between a collinear active quark (or gluon) state in the scattering amplitude and a collinear quark (gluon)-gluon composite state in its complex conjugate amplitude. The relevant TMDs and the quark-gluon correlation functions are closely related to each other. For example, the first $k_\perp$-moment of the
two well-known naive-time-reversal-odd TMDs, the Sivers function $f_{1T}^\perp(x, k_\perp^2)$ \cite{Siv90} and the Boer-Mulders function $h_1^{\perp}(x, k_\perp^2)$ \cite{boermulders}, are equal to the twist-three quark-gluon correlation functions $T_{q,F}(x, x)$ and $T_{q,F}^{(\sigma)}(x, x)$, respectively.  These two correlation functions are defined as \cite{qiu}
\ben
T_{q, F}(x, x)&=&\int\frac{dy_1^- dy_2^-}{4\pi}e^{ixP^+y_1^-}
\langle P,s_T|\bar{\psi}_q(0)\gamma^+\left[ \epsilon^{s_T\alpha n\bar{n}}F_\alpha^{~ +}(y_2^-)\right] \psi_q(y_1^-)|P,s_T\rangle\, ,
\label{Tq}
\\
T_{q, F}^{(\sigma)}(x, x)&=&\int\frac{dy_1^- dy_2^-}{4\pi}e^{ixP^+y_1^-}
\frac{1}{2}\sum_{s_T} \langle P,s_T|\bar{\psi}_q(0)\left[ \sigma^{\alpha +} F_\alpha^{~ +}(y_2^-)\right] \psi_q(y_1^-)|P,s_T\rangle \, ,
\label{Tqs}
\een
where the gauge links between field operators are suppressed and $\epsilon^{0123}=1$ is used. The TMD factorization approach is more suitable for studying scattering processes with two very different momentum transfers, $Q_1\gg Q_2 \gtrsim \Lambda_{\rm QCD}$, where the $Q_2$ is sensitive to the active parton's transverse momentum, while the collinear factorization approach is more relevant for studying scattering cross sections with all observed momentum transfers hard and comparable: $Q_i\sim Q\gg \Lambda_{\rm QCD}$. Although the two approaches each have their own kinematic domain of validity, they are consistent with each other in the regime where they both apply \cite{unify}. 

Both factorization approaches necessarily introduce a factorization scale, $\mu\gg\Lambda_{\rm QCD}$, to separate the calculable short-distance perturbative dynamics from the long-distance nonperturbative physics of the observed cross sections or the asymmetries.  Since the physical observables, the cross sections or the asymmetries, are independent of the choice of the factorization scale, the scale dependence of the nonperturbative distributions \cite{Aybat:2011ge, Kang:2011mr, Kang:2008ey,Zhou:2008mz,Vogelsang:2009pj,Braun:2009mi,Kang:2010xv}, either TMD distributions or twist-3 collinear distributions, must match the scale dependence of corresponding  perturbative hard parts.  That is, the factorization scale dependence of the nonperturbative distributions is perturbatively calculable and is a prediction of QCD perturbation theory when $\mu\gg\Lambda_{\rm QCD}$. For example, the scale dependence of the leading power parton distributions obeys DGLAP evolution equations whose evolution kernels are perturbatively calculable, and has been very successfully tested when the scale varies from a few GeV to the hundreds of GeV. 

Recently the scale dependence of the twist-three quark-gluon correlation functions $T_{q, F}(x, x)$ and $T_{q, F}^{(\sigma)}(x, x)$ have also been studied by several groups \cite{Kang:2008ey,Zhou:2008mz,Vogelsang:2009pj,Braun:2009mi}. However, there are discrepancies between these results, particularly for the evolution of  $T_{q, F}(x, x)$ (also often refer to as Efremov-Teryaev-Qiu-Sterman (ETQS) function): the result in Ref.~\cite{Braun:2009mi} is different from those in Refs.~\cite{Kang:2008ey,Zhou:2008mz,Vogelsang:2009pj} by two extra terms. The purpose of our paper is to identify and resolve these discrepancies, and in addition we also derive the evolution equations for the other quark-gluon correlation function $T_{q, F}^{(\sigma)}(x, x)$.

\bef
\psfig{file=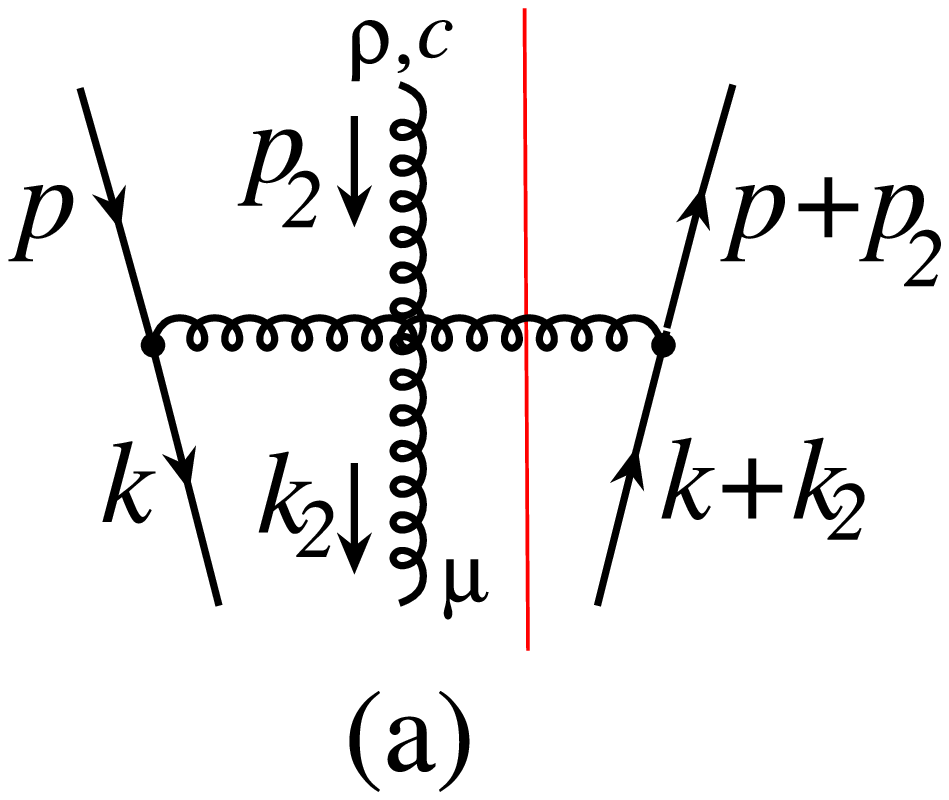, width=1.3in}
\psfig{file=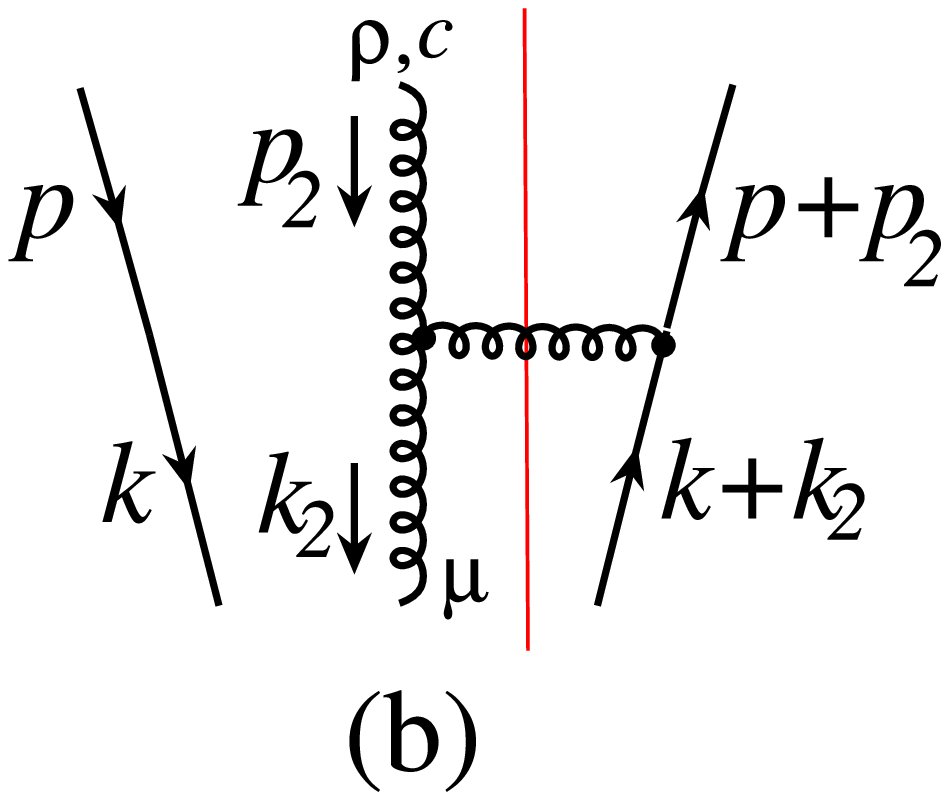, width=1.3in}
\psfig{file=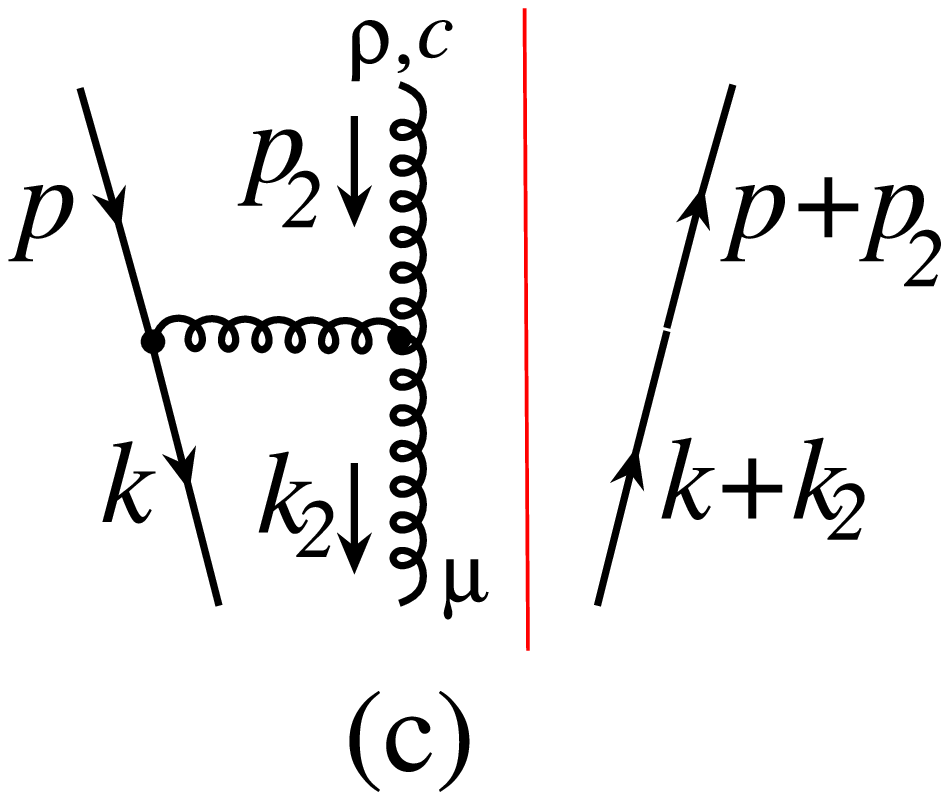, width=1.3in}
\psfig{file=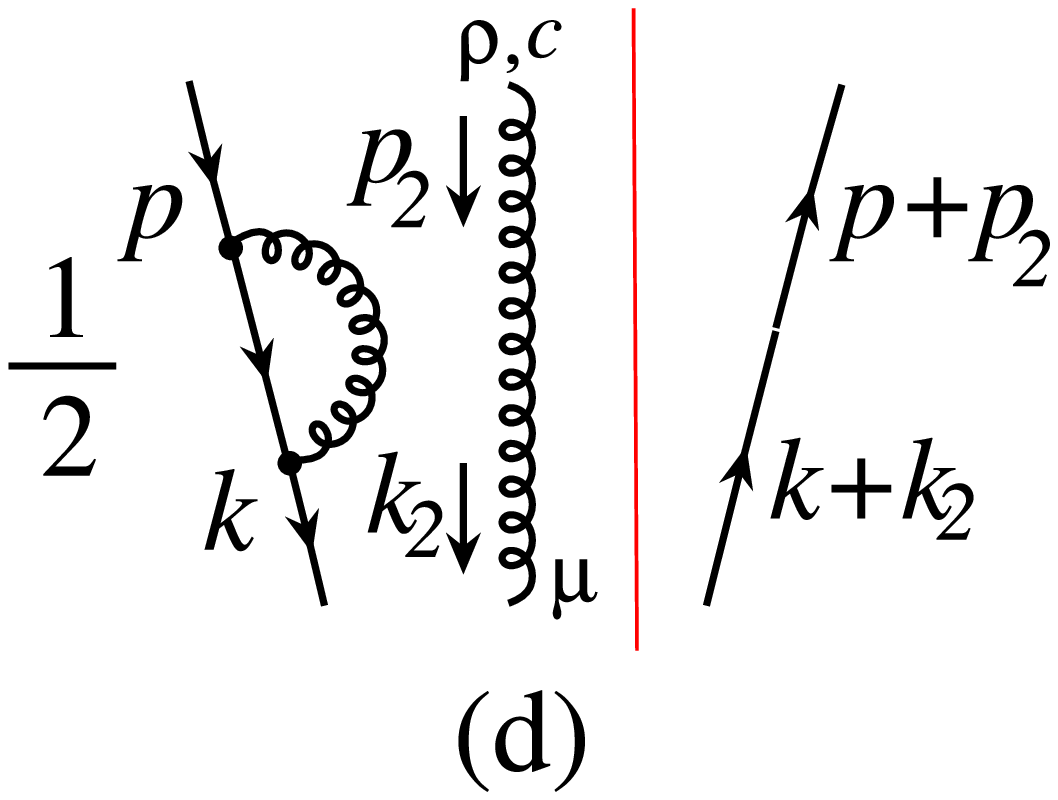, width=1.3in}
\psfig{file=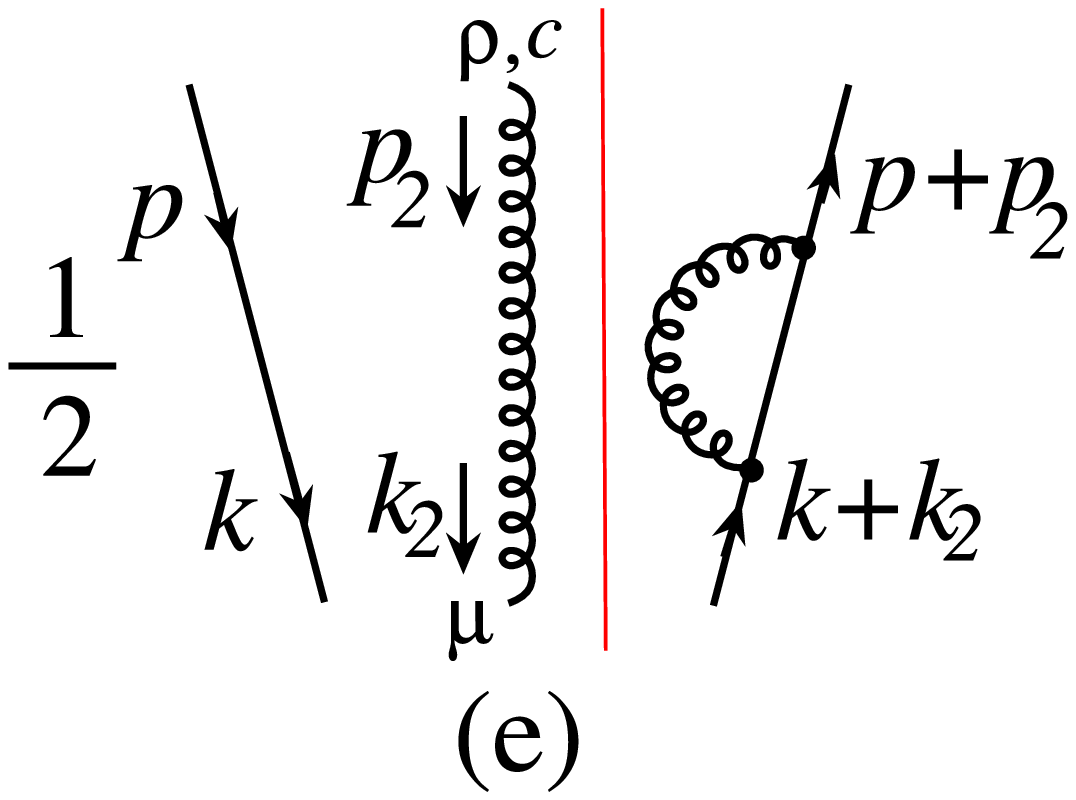, width=1.3in}
\caption{Feynman diagrams contribute to the leading order evolution kernel of quark-gluon correlation functions.}
\label{lo}
\eef 
The evolution or the factorization scale dependence of these twist-3 parton distributions is an immediate consequence of the QCD factorization formalism for physical observables.   Since these twist-3 parton distributions are process independent, the perturbative evolution kernels are universal, although they could be derived in many different ways.  In Ref.~\cite{Kang:2008ey}, we presented a derivation for the evolution equations of $T_{q, F}(x, x)$, as well as other twist-3 correlation functions relevant to single transverse spin asymmetries, from the perturbative variation of these functions. By calculating the leading order Feynman diagrams \cite{Kang:2008ey}, we obtained finite contributions from those diagrams in Fig.~\ref{lo} and the following evolution equation,
\ben
\frac{\partial {T}_{q,F}(x,x,\mu)}{\partial{\ln \mu^2}}
&=&
\frac{\alpha_s}{2\pi}
\int_x^1\frac{d\xi}{\xi}
\bigg\{
P_{qq}(z)\, {T}_{q,F}(\xi,\xi,\mu)
\nnu
&\ & \hskip 0.6in
+\frac{N_c}{2}
\left[\frac{(1+z){T}_{q,F}(\xi,x,\mu)
     -(1+z^2){T}_{q,F}(\xi,\xi,\mu)}{1-z}
     +{T}_{\Delta q,F}(x,\xi,\mu)\right]\bigg\},
\label{old}
\een
where $z=x/\xi$ and $P_{qq}(z)$ is the splitting kernel for unpolarized quark distribution function given by
\ben
P_{qq}(z) = C_F\left[\frac{1+z^2}{(1-z)_+}+\frac{3}{2}\delta(1-z) \right],
\een
and the quark-gluon correlation function ${T}_{\Delta q,F}(x_1, x_2, \mu)$ is given by \cite{Kang:2008ey}
\ben
T_{\Delta q,F}(x_1, x_2)
=\int\frac{dy_1^- dy_2^-}{4\pi}\,
e^{i x_1 P^+y_1^-} e^{i (x_2-x_1) P^+y_2^-}
\langle P,s_T|\bar{\psi}_q(0)\,
\gamma^+\gamma^5
\left[i\, s_T^\alpha \, F_\alpha^{~ +}(y_2^-)\right] 
\psi_q(y_1^-)|P,s_T\rangle.
\label{TDqt}
\een
The results derived in Refs.~\cite{Zhou:2008mz,Vogelsang:2009pj} are consistent with ours.  But, the evolution equation derived later by Braun, Manashov, and Pirnay in Ref.~\cite{Braun:2009mi} is slightly different,
\ben
\frac{\partial {T}_{q,F}(x,x,\mu)}{\partial{\ln \mu^2}}
&=&
\frac{\alpha_s}{2\pi}
\int_x^1\frac{d\xi}{\xi}
\bigg\{
P_{qq}(z)\, {T}_{q,F}(\xi,\xi,\mu)
\nnu
&\ & \hskip 0.6in
+\frac{N_c}{2}
\left[\frac{(1+z){T}_{q,F}(\xi,x,\mu)
     -(1+z^2){T}_{q,F}(\xi,\xi,\mu)}{1-z}
     -{T}_{\Delta q,F}(x,\xi,\mu)\right]
\nnu
&\ & \hskip 0.6in
-N_c \, \delta(1-z)\, {T}_{q,F}(x, x, \mu)
+\frac{1}{2N_c}\left[ (1-2 z) {T}_{q,F}(x, x-\xi, \mu) - {T}_{\Delta q,F}(x, x-\xi, \mu) \right]\bigg\}.
\label{braun}
\een

Comparing Eqs.~(\ref{old}) and (\ref{braun}), it is clear that two results differ by two contributions listed in the third line in Eq.~(\ref{braun}). In addition, there is a sign difference in front of the ${T}_{\Delta q,F}$ distribution in the second line. This sign difference is due to a fact that two groups used a different sign convention for anti-symmetric tensor $\epsilon^{\mu\nu\alpha\beta}$:  we chose $\epsilon^{0123}=1$, while Braun-Manashov-Pirnay used $\epsilon_{0123}=1$ implying $\epsilon^{0123}=-1$\footnote{This was also speculated in Ref.~\cite{Braun:2009mi}.}.  We also noticed that Ref.~\cite{Ma:2011nd} used the same convention as that in our paper, thus they obtain the same sign for the ${T}_{\Delta q,F}$ term.

\bef
\psfig{file=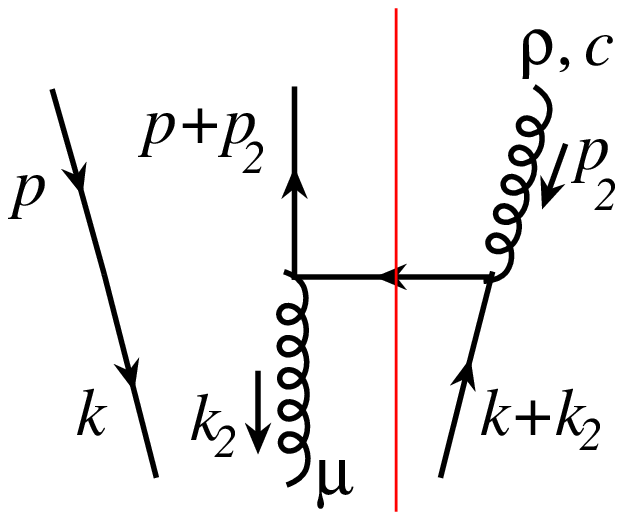, width=1.3in}
\hskip 0.3in
\psfig{file=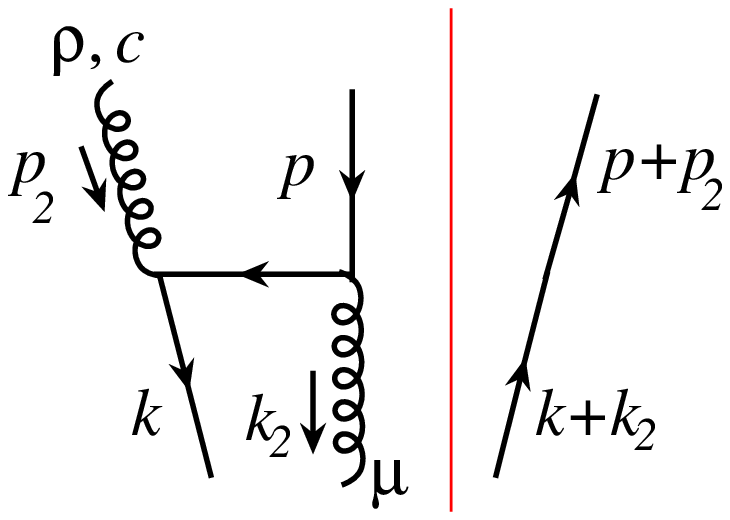, width=1.45in}
\caption{Feynman diagrams contribute to the evolution from the interference of a gluon and a quark-antiquark state (left), and the usual interference of a quark and a quark-gluon state (right).}
\label{extra}
\eef
In the third line in Eq.~(\ref{braun}), the second term $\propto 1/2N_c$  comes from the two Feynman diagrams in Fig.~\ref{extra}. The left diagram in Fig.~\ref{extra} corresponds to the interference between a gluon and a quark-antiquark pair. This diagram was missed in our original calculation in Ref.~\cite{Kang:2008ey}. Once we calculate this diagram, we obtain half of the $1/2N_c$ term in Eq.~(\ref{braun}),
\ben
\left.\frac{\partial {T}_{q,F}(x,x,\mu)}{\partial{\ln \mu^2}}\right|_{\rm Fig.~\ref{extra}(left)} 
= \frac{\alpha_s}{2\pi}
\int_x^1\frac{d\xi}{\xi}
\frac{1}{2N_c}\left(\frac{1}{2}\right)\left[ (1-2 z) {T}_{q,F}(x, x-\xi, \mu) + {T}_{\Delta q,F}(x, x-\xi, \mu) \right],
\een
where again the ${T}_{\Delta q,F}$ term has an overall sign difference due to our convention for $\epsilon^{\mu\nu\alpha\beta}$. The right diagram in Fig.~\ref{extra} is actually Fig.~7(h) in our original paper Ref.~\cite{Kang:2008ey}. This diagram vanishes if the quark on the left of the cut has a positive momentum $p^+=\xi P^+ > 0$, which was assumed in the original paper \cite{Kang:2008ey}. However, the $\xi$ does not have to be larger than 0 as long as  $\xi+\xi_2 > 0$.  By calculating the contribution from the region where $\xi < 0$, we find that it gives exactly the other half of the $1/2N_c$ term in Eq.~(\ref{braun}),
\ben
\left.\frac{\partial {T}_{q,F}(x,x,\mu)}{\partial{\ln \mu^2}}\right|_{\rm Fig.~\ref{extra}(right)} 
= \frac{\alpha_s}{2\pi}
\int_x^1\frac{d\xi}{\xi}
\frac{1}{2N_c}\left(\frac{1}{2}\right)\left[ (1-2 z) {T}_{q,F}(x-\xi, x, \mu) - {T}_{\Delta q,F}(x-\xi, x, \mu) \right].
\een
In other words, adding two diagrams together we have
\ben
\left.\frac{\partial {T}_{q,F}(x,x,\mu)}{\partial{\ln \mu^2}}\right|_{\rm Fig.~\ref{extra}(left+right)} 
= \frac{\alpha_s}{2\pi}
\int_x^1\frac{d\xi}{\xi}
\frac{1}{2N_c}\left[ (1-2 z) {T}_{q,F}(x, x-\xi, \mu) + {T}_{\Delta q,F}(x, x-\xi, \mu) \right].
\een

The other term in the third line in Eq.~(\ref{braun}), $-N_c T_{q,F}(x, x)$, was missed in our original paper \cite{Kang:2008ey}. The error was caused by a subtlety in taking the limit $x_2\to 0$ when we evaluate the integration $\int dx_2 \delta(x_2) x_2 F(x_2)=\lim_{x_2\to 0} x_2 F(x_2)$ to get the gluonic-pole matrix element.   The limit, $\lim_{x_2\to 0} x_2 F(x_2)$, would vanish if the function $F(x_2)$ is finite as $x_2\to 0$, which is unfortunately not always true in our calculation.  We find that Fig.~\ref{lo}(b) and (c) have additional contributions to the evolution as,
\ben
\left.\frac{\partial {T}_{q,F}(x,x+x_2,\mu)}{\partial{\ln \mu^2}}\right|_{\rm Fig.~\ref{lo}(b)-additional} & = & \frac{N_c}{2} \int_{x_2}^{1-x} 
d\xi_2 \,T_{q,F}(x, x+\xi_2, \mu) \left[- \frac{x_2}{\xi_2^2}\right] + \cdots,
\label{7b}
\\
\left.\frac{\partial {T}_{q,F}(x,x+x_2,\mu)}{\partial{\ln \mu^2}}\right|_{\rm Fig.~\ref{lo}(c)-additional} & = & \frac{N_c}{2} \int_{x+x_2-1}^{x_2} 
d\xi_2 \,T_{q,F}(x+x_2-\xi_2, x+x_2, \mu) \left[\frac{x_2}{\xi_2^2}\right] + \cdots \, .
\label{7c}
\een 
Here the ``$\cdots$'' includes any regular terms which vanish safely when we take $x_2\to 0$. The subtlety is caused by the fact that the integration over $d\xi_2$ in Eqs.~(\ref{7b}) and (\ref{7c}) is singular as $x_2\to 0$.  To evaluate the $d\xi_2$ integration, we first expand $T_{q,F}(x, x+\xi_2, \mu)$ around $\xi_2=x_2$ in Eq.~(\ref{7b}),
\ben
T_{q,F}(x, x+\xi_2, \mu) = T_{q,F}(x, x+x_2, \mu) + \frac{\partial}{\partial \xi_2}T_{q,F}(x, x+\xi_2, \mu)|_{\xi_2\to x_2} (\xi_2 - x_2) + \cdots \, .
\label{expansion}
\een
The integration in Eq.~(\ref{7b}) with the first term of the expansion in Eq.~(\ref{expansion}) gives 
\ben
\frac{N_c}{2} \int_{x_2}^{1-x} 
d\xi_2 \,T_{q,F}(x, x+x_2, \mu) \left[- \frac{x_2}{\xi_2^2}\right] = \frac{N_c}{2}\,T_{q,F}(x, x+x_2, \mu)
\left[\frac{x_2}{\xi_2}\right]^{1-x}_{x_2} = \frac{N_c}{2}\,T_{q,F}(x, x+x_2, \mu)
\left[\frac{x_2}{1-x} - 1\right],
\een
which goes to $-\frac{N_c}{2} T_{q,F}(x, x, \mu)$ at the limit $x_2\to 0$. If one assumes $T_{q,F}(x, x+\xi_2, \mu)$ is a smooth (regular) function, we find that the higher order terms in the expansion in Eq.~(\ref{expansion}) do not contribute to the evolution in Eq.~(\ref{7b}) when $x_2\to 0$. So Fig.~\ref{lo}(b) gives us an additional contribution $-\frac{N_c}{2} T_{q,F}(x, x, \mu)$. Similarly we find exactly the same contribution from Eq.~(\ref{7c}). Adding them together, we have
\ben
\left.\frac{\partial {T}_{q,F}(x,x,\mu)}{\partial{\ln \mu^2}}\right|_{\rm Fig.~\ref{lo}(b+c)-additional} 
 = - N_c \,T_{q,F}(x, x, \mu),
\een
which is exactly what was missed in our original paper \cite{Kang:2008ey}. 

We now have a complete agreement with the Braun-Manashov-Pirnay result. In other words, in our $\epsilon^{0123}=1$ convention, we have the flavor nonsinglet evolution equation for ${T}_{q,F}(x,x,\mu)$ as
\ben
\frac{\partial {T}_{q,F}(x,x,\mu)}{\partial{\ln \mu^2}}
&=&
\frac{\alpha_s}{2\pi}
\int_x^1\frac{d\xi}{\xi}
\bigg\{
P_{qq}(z)\, {T}_{q,F}(\xi,\xi,\mu)
\nnu
&\ & \hskip 0.6in
+\frac{N_c}{2}
\left[\frac{1+z^2}{1-z}\left({T}_{q,F}(\xi,x,\mu)
    -  {T}_{q,F}(\xi,\xi,\mu)\right)
     + z \, {T}_{q,F}(\xi,x,\mu)
     +{T}_{\Delta q,F}(x,\xi,\mu)\right]
\nnu
&\ & \hskip 0.6in
-N_c \, \delta(1-z)\, {T}_{q,F}(x, x, \mu)
+\frac{1}{2N_c}\left[ (1-2 z) {T}_{q,F}(x, x-\xi, \mu) + {T}_{\Delta q,F}(x, x-\xi, \mu) \right]\bigg\}.
\label{new}
\een
Similarly, our results for flavor singlet evolution are also now consistent with the Braun-Manashov-Pirnay result.

Using the same techniques, we could also derive the evolution equation for the other twist-three quark-gluon correlation function ${T}_{q,F}^{(\sigma)}(x,x,\mu)$. The calculation is straightforward, and the result is
\ben
\frac{\partial {T}_{q,F}^{(\sigma)}(x,x,\mu)}{\partial{\ln \mu^2}}
&=&
\frac{\alpha_s}{2\pi}
\int_x^1\frac{d\xi}{\xi}
\bigg\{
\Delta_{T}P_{qq}(z)\, {T}_{q,F}^{(\sigma)}(\xi,\xi,\mu)
+\frac{N_c}{2}
\left[\frac{2 \, {T}_{q,F}^{(\sigma)}(\xi,x,\mu)
     -2 z\, {T}_{q,F}^{(\sigma)}(\xi,\xi,\mu)}{1-z}\right]
\nnu
&\ & \hskip 0.6in
-N_c \, \delta(1-z)\, {T}_{q,F}^{(\sigma)}(x, x, \mu)
+\frac{1}{2N_c} \left[2 (1-z) {T}_{q,F}^{(\sigma)}(x, x-\xi, \mu)\right] \bigg\},
\label{sigma}
\een
where $\Delta_{T}P_{qq}(z)$ is the splitting kernel for the quark transversity given by
\ben
\Delta_{T}P_{qq}(z) = C_F\left[\frac{2\, z}{(1-z)_+}+\frac{3}{2}\delta(1-z) \right].
\een
This evolution equation was first derived in Ref.~\cite{Zhou:2008mz}, which contains only the first line in Eq.~(\ref{sigma}). The first term  in the second line, $-N_c\, {T}_{q,F}^{(\sigma)}(x, x, \mu)$, has exactly the same origin as those in Eqs.~(\ref{7b}) and (\ref{7c}) from calculating diagrams in Figs.~\ref{lo}(b) and (c) with a caution of taking the limit $x_2\to 0$.  The second term, $\propto 1/2N_c$,  is again due to the fact that the Feynman diagrams in Fig.~\ref{extra} were not included in the calculation of Ref.~\cite{Zhou:2008mz}.

In summary, we have rederived the evolution equations for both ${T}_{q,F}(x, x, \mu)$ and ${T}_{q,F}^{(\sigma)}(x, x, \mu)$. We resolved the discrepancies in the literature for the evolution of ETQS function 
${T}_{q,F}(x, x, \mu)$. We understand that such discrepancies were also resolved by the other two groups \cite{Schafer:2012ra, WV} through careful reexaminations of their original derivations in Refs.~\cite{Zhou:2008mz} and \cite{Vogelsang:2009pj}, also in Ref.~\cite{Ma} from a different approach. Using the same techniques developed in the current paper, we updated the calculation for the evolution of ${T}_{q,F}^{(\sigma)}(x, x, \mu)$ and found two additional contributions which are missing in the literature. These results will have important consequences, e.g., in the study of QCD resummation for the spin-dependent observables \cite{Kang:2011mr}.

We thank J.~P.~Ma, F.~Yuan and J.~Zhou for useful communications. This work was supported in part by the US Department of Energy, Office of Science, under Contract No.~DE-AC52-06NA25396 and DE-AC02-98CH10886.


\end{document}